\documentclass[11pt,a4paper]{article}
\usepackage{psfig}
\begin{document}
	\title{Folding pathways of a helix-turn-helix model protein}

\author{Daniel Hoffmann\footnote{Present address:
German National Research Center for Information Technology,
GMD-SCAI,
Schloss Birlinghoven,
D-53754 Sankt Augustin, Germany; e-mail daniel.hoffmann@gmd.de; URL http://www.gmd.de/SCAI/people/hoffmann.html}
\hspace{1ex}and 
Ernst-Walter Knapp\\
Freie Universit\"at Berlin,\\
Fachbereich Chemie, Institut f\"ur Kristallographie,\\
Takustr.~6, D-14195 Berlin, Germany}
\date{J. Phys. Chem., in press}
\maketitle

\begin{abstract}
\baselineskip0.8cm
A small model polypeptide represented in atomic detail is folded using
Monte Carlo dynamics. The polypeptide is designed to have a native
conformation similar to the central part of the helix-turn-helix
protein ROP. Starting from a $\beta$-strand conformation
or two different loop conformations of the
protein glutamine synthetase, six trajectories are generated using
the so-called window move in dihedral angle space. This move changes
conformations locally and leads to realistic, quasi-continuously
evolving trajectories. Four of the six
trajectories end in stable native-like conformations. Their folding
pathways show a fast initial development of a helix-bend-helix motif,
followed by a dynamic behaviour predicted by the diffusion-collision
model of Karplus and Weaver. The phenomenology of the pathways is
consistent with experimental results.  
\end{abstract}

\newpage

	\section*{Introduction}
One of the most intriguing problems in molecular biology is the
decryption of the protein folding code. There is a wealth of
experimental results providing insights into kinetics and
thermodynamics of the folding process \cite{Fer1995,Mir1996}.  They
point to a delicate interplay of hydrophobic and electrostatic
interactions which guide the formation of secondary and tertiary
structures.  Computer experiments on protein folding using Monte Carlo
(MC) methods with simplified lattice models also contributed very much
to our understanding of the principles of protein folding
\cite{Dil1995,Kar1995}. With these simplified protein models general
aspects of protein folding can be studied. However they are not
designed to reflect the behavior of a particular existing protein at
the atomic level of description. The latter is the domain of
conventional molecular dynamics (MD)
\cite{Bro1983,Kar1986,McC1987,Gun1988}. Unfortunately conventional MD
cannot be used to address the protein folding problem for the time
being.  The reason for this becomes clear if we consider that the the
typical time propagation step in MD is 1~fs, and that the CPU-time per
evaluation of the energy function is of the order of 1~s. Hence
typical folding processes lasting 1~ms to 1~s would require CPU-times
of 10$^{12}$~s to 10$^{15}$~s.

Off-lattice MC dynamics is an alternative that can extend the time
range accessible to simulation into the folding regime
\cite{Lev1975}. We have proposed a MC method
\cite{Kna1992,Hoffmann1996a,Hoffmann1996b} which combines a detailed
protein model, with dihedral angles as continuous degrees of freedom,
and an efficient algorithm for the generation of new
conformations. This so-called window algorithm simulates the evolution
of the polypeptide conformation by series of local conformational
changes, each one restricted to a window, i. e. a randomly selected
short stretch of polypeptide backbone. As has been shown earlier
\cite{Hoffmann1996b} the local move, with its cooperative changes of
dihedral angles in the window, performs far better than a simple MC
move where torsion angles of the polypeptide backbone are changed
independently and thus global conformational changes are
generated. Furthermore window moves simulate the dynamical time
evolution at least two orders of magnitude faster than MD
\cite{Kna1992a,Hoffmann1996b}.

In this paper we study MC trajectories for a small model protein in
detail. These trajectories provide an atomistic view of possible
mechanisms in the protein folding process and offer interpretations
for experimental results like the fast formation of secondary
structure \cite{Kuw1987,Udg1988}, the existence of transient
non-native conformations \cite{Dys1992a,Log1994,Zha1995}, or the
existence of multiple pathways \cite{Rad1992}. It turns out that the
time evolution observed in the trajectories is in good agreement with
predictions made by the diffusion-collision model (DCM) of Karplus and
Weaver \cite{Kar1976}. This model allows a quantitative description
of the folding mechanism first proposed by Ptitsyn and Rashin
\cite{Pti1973}, and later supported by Kim and Baldwin
\cite{Kim1982}, who coined the name ``framework model''. The DCM
essentially states that the first step in folding is the formation of
micro\-domains, e.~g. $\alpha$-helices, which diffuse relatively to each
other and eventually collide and coalesce with a certain
probability. In this way new and larger domains are formed, which again
collide and coalesce, etc. In this sense the formation of an
$\alpha$-helical hairpin can be understood as an elementary event of
the DCM.

	\section*{The model system}
The simulation of protein folding with a detailed model demands great
amounts of CPU-time. Therefore it is important to choose a protein
that is as simple as possible. It nevertheless should have features
typical of proteins, i. e. a native state with secondary and tertiary
structure. These structural elements are stabilized by hydrogen bonds
and hydrophobic interactions. Hence, the model should consider
corresponding energy terms. ROP is a simple protein that has secondary
structure and tertiary contacts \cite{Ban1987}. It forms an
$\alpha$-helix-turn-$\alpha$-helix, that is a so-called
$\alpha$-helical hairpin motif. The central part of ROP, consisting of
the 26 residues from 18 to 43, is used as a template for the
construction of a model polypeptide. Three types of amino acids are
used. The five residues in the central turn are replaced by
glycines (G) which are known to have a high propensity to form
turns. The residues responsible for interhelical contacts are assigned
to residues of type X that differ from alanines only in the increased
attraction between C$_\beta$-atoms of X residues, mimicking a
hydrophobic interaction. This attraction is modeled by a Lennard-Jones
potential with a well-depth of 2.0~kcal/mol, a value that is motivated
by the free energy changes for transfer of typical hydrophobic amino
acids from a non-polar to a polar solvent \cite{Kau1959,Cho1974}. All
remaining residues are replaced by alanines (A). Alanines are often
found in $\alpha$-helical secondary structure of proteins, and in MD
simulations of polyalanine $\alpha$-helices are formed {\em in vacuo}
\cite{Bro1989,Dag1991}. The whole sequence of the model polypeptide
reads AXAAXAAAXXGGGGGXXAAAXAAAXA. The terminal alanines are blocked
with neutral acetamide and N-methyl-amide groups to avoid strong
Coulombic interactions. Since each of the sidechains of A and X is
represented by a single, so-called extended $C_\beta$-atom
\cite{Bro1983}, there are no torsional degrees of freedom in the
sidechains. The similarity of the model
polypeptide with the original ROP is merely a
structural one, insofar 
as it has a high propensity to form a $\alpha$-helical hairpin. With
respect to other properties there may be significant differences between
model polypeptide and ROP, e.g. the ROP monomer is not stable in
aqueous solution \cite{Steif1993}.

The bond lengths and bond angles of the model polypeptide, as well as
the dihedral angles
$\omega_i(C_{\alpha,i}C_iN_{i+1}C_{\alpha,i+1})$ are fixed to
equilibrium values provided by the parameter set of the MD programme
CHARMM22 \cite{Bro1983}. Thus the only remaining degrees of freedom
of the model polypeptide are the dihedral angles
$\phi(C_{i-1}N_iC_{\alpha,i}C_i)$ and $\psi
(N_iC_{\alpha,i}C_iN_{i+1})$ in the backbone. The force field is
adopted from the MD programme CHARMM, with specific changes to
compensate the greater rigidity of the polypeptide model due to the
fixed bond lengths and bond angles. This compensation is achieved by
replacing the explicit atom pair interactions between sequentially
neighbouring amide planes by an effective two-dimensional
($\phi,\psi$) torsion potential which implicitely considers the
flexibility of the rigidified degrees of freedom. This torsion
potential is obtained once by constrained energy minimization of
dipeptides with fixed values of $\phi$ and $\psi$ but all other
degrees of freedom unconstrained, as described by Brooks {\em et al.}
in Appendix 2 of Ref. \cite{Bro1983}. Apart from the replacement of
the respective atom pair interactions by the torsion potential, the
energy function is that of CHARMM \cite{Bro1983}.

For each move a window is placed randomly on the polypeptide, the
conformation is changed in the window, and the energy of the new
conformation is evaluated. The difference of energy between the new
and the preceding conformation is then used in the criterion of
Metropolis {\em et al.}  \cite{Met1953} to decide whether the MC
move is accepted or rejected. Conformations generated by this
procedure represent a canonical ensemble at the given temperature
$T$. In the present case only windows containing three peptide planes
are used because there the acceptance probability is particularly
favorable with values ranging from 0.3 to 0.4.

In a first step the rigidified backbone of the model polypeptide was
fitted to the considered central part of the x-ray structure of ROP by
minimizing the root mean square deviation (RMSD) with respect to the
backbone atoms. In this ROP-fitted conformation the model polypeptide
possesses five interhelical X--X pairs with C$_\beta$-C$_\beta$
distances of less than 6~{\AA} and a well developed system of
$\alpha$-helical hydrogen bonds in each of the two helices. There are
no strains in this conformation as can be concluded from energy
minimization, that is Metropolis MC at $T=0$~K, where the conformation
changes by less than 1~{\AA} RMSD and the energy drops from
$-1140$~kcal/mol to $-1200$~kcal/mol. Thus the designed amino acid
sequence has a native conformation close to that of the ROP-fitted
conformation. This assumption is supported further by eight simulated
annealing simulations with initial temperatures of 1000--3000~K
starting from the ROP-fitted conformation of the model polypeptide. In
these simulations the conformation of lowest energy
($-$1215~kcal/mol), the ``reference structure'', had a RMSD of
1.35~{\AA} to the ROP-fitted conformation.

It has been shown elsewhere \cite{Hoffmann1996b} that window moves are
able to simulate the folding of this model polypeptide. Here we
analyse the folding process in more detail. Six
trajectories were generated at $T=450$~K, a temperature low enough for
the designed native conformation being stable, and high enough to
facilitate folding within a reasonable short amount of CPU-time. The
elevated temperature can compensate for the missing aqueous solvent,
whose presence would weaken intra-polypeptide interactions like
hydrogen bonding.  Furthermore isomerization barriers are still
somewhat to high compared with those encountered in MD simulations
despite the adaptation of the energy function to the rigidified
polypeptide model. Thus $T=450$~K corresponds to a lower temperature
for MD simulations in aqueous solution.

	\section*{Results and discussion}
Six trajectories were generated. Four of them
($\beta$-trajectories) started from a $\beta$-strand conformation
($\phi=-120^\circ, \psi=120^\circ$). This conformation is quite
elongated, and thus far away from the ``native'' helical hairpin
structure, but unlike the fully extended conformation, for the
$\beta$-strand the interactions of neighbouring peptide planes are in
a local minimum. Hence the $\beta$-strand seems to be a reasonable
model for a denaturated state. In order to investigate the sensitivity
of the results to the initial conditions, the remaining two
trajectories (loop-trajectories) started from two other conformations,
namely the one of loop Glu~13 -- Asn~39 and of loop Lys~163 -- Gln~189
(Fig.~\ref{fig:loopStarts}), respectively, of chain F in the protein
complex glutamine synthetase (PDB code 2gls)
\cite{Yamashita1989}. These two loops have completely different
structures and are devoid of helical turns. Hence there is no bias
towards the $\alpha$-helical hairpin. Five
out of the six trajectories end up in
helix-turn-helix conformations which after minimization have lower
energies than the reference structure introduced in the previous
section. The two loop-trajectories and two of
the $\beta$-trajectories are well equilibrated after about $10^6$ MC
scans of window moves (in a MC scan the window is placed randomly on
the polypeptide backbone as many times as there are possible window
positions). In the end they show only modest fluctuations of the
energy and of structural quantities like the radius of gyration. These
four trajectories deliver conformations close
to the reference structure (all-atom root mean square deviations 2.3,
2.7, 3.2, 3.8 {\AA}, respectively), and the number of interhelical
X--X pairs with C$_\beta$-C$_\beta$ distance smaller than 6~{\AA} is
five to six, which is identical to that of the reference
structure. The mean energies in the
equilibrated parts of the four trajectories lie at about
-1190~kcal/mol. The other two
$\beta$-trajectories lead to a single long helix, and a
helix-turn-helix motif with some left handed helical turns,
respectively. The single helix shows fraying at the termini and also
bending dynamics, but develops no stable loop between two helices. Due
to the missing X--X interactions the mean energy (-1175~kcal/mol) of
the trajectory is significantly higher than of the other
trajectories. The same holds true for the structure with the left
handed helix turns, which prevent the formation of all possible
interhelical X--X pairs. These two trajectories hence represent
different free energy minima above the minimum corresponding to the
``native'' helical hairpin. They could reach this lower minimum by
breaking some helical hydrogen bonds in the case of the long helix, or
by expansion and refolding with the correct righthanded helical turns
in the case of the other trajectory. Both trajectories explore
conformations towards these barriers but do not cross them during the
simulation of $2\cdot 10^6$ MC scans.  In the following we focus
mainly on the folding paths observed in the two equilibrated
$\beta$-trajectories, which we call trajectory (1) and (2),
respectively. These two trajectories are representative for those four
trajectories which converge into $\alpha$-helical hairpins. If not
mentioned otherwise the described observations refer to both
trajectories.

Within the first few tens of thousands of MC scans the elongated
$\beta$-strand relaxes into a mainly $\alpha$-helical conformation
(Fig.~\ref{fig:sdg}(b) and Fig.~\ref{fig:sdg}(c)). The
initiations of the helices take place almost simultaneously at
different locations of the polypeptide, but preferentially near the
termini. The only exception is the C-terminal helix in trajectory (1)
which begins to grow at residues 17, 18 and 19. Each helix grows until
it reaches the nearest terminus, or until it is stopped by a multiple turn
structure near the center of the polypeptide at one end of the stretch
of glycines. At this stage the turn structure consists mainly of X and
A residues, and most of the glycines are incorporated into one of the
two helices. This is particularly noteworthy because usually glycines
act as helix breakers. However there are experiments indicating that
non-native conformations can have appreciably populated 
non-native secondary structures \cite{Log1994,Zha1995}. In this
context it is interesting to note that peptide fragments of
myohemerythrin \cite{Dys1992a} and plastocyanin \cite{Dys1992b}
which include the loop and turn regions of the native proteins clearly
show $\alpha$-helix content. 

The turn region functions as a buffer between the N- and C-terminal
helices, preventing them from unification. The helix growth is
accompanied by a sharp drop in energy (Figs.~\ref{fig:combi1}(a) and
\ref{fig:combi2}(a)) from $-1008$~kcal/mol for the initial
$\beta$-strand conformation to about $-1175$~kcal/mol, mainly
originating from the formation of helical hydrogen bonds and other
attractive sequence local interactions. The increasing helix content
is also visible in the contraction from 25~{\AA} to 10~{\AA} radius of
gyration $R_{gyr}$, or 82~{\AA} to 32~{\AA} end-end-distance
$R_{ends}$ (Figs.~\ref{fig:combi1}(c) and \ref{fig:combi2}(c)). The
observed fast formation of secondary structure as a first step in the
folding process is consistent with experimental findings for a variety
of proteins, like cytochrome c and $\beta$-lactoglobulin
\cite{Kuw1987}, ribonuclease A \cite{Udg1988}, lysozyme
\cite{Rad1992}, or {\em Escherichia coli} trp aporepressor
\cite{Man1993}.

The phase of helix formation ends after $10^5$ MC scans and can be
clearly distinguished from the following second phase in which the
helix content remains essentially constant but $R_{ends}$ fluctuates
considerably. In other words the two $\alpha$-helices are diffusing
relatively to each other as quasi rigid entities with only the
interhelical angle changing randomly (Figs.~\ref{fig:combi1}(d) and
\ref{fig:combi2}(d)). In trajectory (1) this diffusion leads to a near
coalescence of the two helices after about \linebreak $3\cdot 10^5$ MC
scans where $R_{ends}$ and $R_{gyr}$ drop sharply. Simultaneously the
number of pairs of non-neighbouring hydrophobic X residues with
C$_\beta$-C$_\beta$ distances less than 6~{\AA} increases transiently
from one or two to six. At the same time the short N-terminal helix
unravels almost completely (Fig.~\ref{fig:sdg}(b)). During this
near-coalescence the energy trace shows no dramatic change. The new
more compact conformation is unstable and decays into a more elongated
one, while the short N-terminal helix recovers. The diffusion of the
angle between the two helices lasts for about $3\cdot 10^5$ and $10^5$
MC scans in trajectory (1) and (2), respectively. The shorter
diffusion phase in trajectory (2) may account for the lack of
near-coalescence events like that observed in trajectory (1). For both
trajectories the energy has a value of $-1160\pm10$~kcal/mol during
the helix angle diffusion.

It is notable that trajectories (1) and (2), although being formally
very similar and undistinguishable if for example only the development
of the helix content would be monitored, represent two different
folding pathways. In (1) the turn forms in the N-terminal half of the
sequence (Fig.~\ref{fig:sdg}(b)), whereas in (2) it forms in the
C-terminal half (Fig.~\ref{fig:sdg}(c)). Thus in (1) a short
N-terminal helix coalesces with a longer C-terminal one, while in (2)
a short C-terminal helix joins a longer N-terminal one
(Figs.~\ref{fig:creeping1} and \ref{fig:creeping2}). The existence of
multiple folding pathways has also been suggested to explain
experiments on ribonuclease \cite{Bal1990} and lysozyme
\cite{Rad1992}.

The period of relative diffusion of the interhelical angle is
terminated by the coalescence of the two helices. The coalescence is
initiated by the formation of a cluster of hydrophobic X residues at
the shorter helix, inducing a sharp reverse turn and bending the
shorter helix towards the longer one. Then this bent conformation
leads to the formation of the first interhelical hydrophobic X-X
pairs. The number of X-X pairs is still small, indicating the
non-native character of the conformations in this phase. The
non-native character can also be seen from the fact that shortly after
coalescence the glycines in the center of the sequence are still part
of one of the helices and hence the $\alpha$-hairpin is quite
asymmetric with a shorter and a longer helix (trajectory (1) after
460000 MC scans, Fig.~\ref{fig:creeping1}, and trajectory (2) after
160000 MC scans, Fig.~\ref{fig:creeping2}).

Now a new type of movement can be observed. The respective shorter
helix begins to creep along the longer one. This movement is driven by
the attraction of X residues in interhelical X-X pairs and leads to
the formation of an increasing number of X-X pairs. The creeping
generates a pull which is transmitted onto the longer helix through
the connecting reverse turn. Eventually this pull forces the glycines
out of the helical turn into the reverse turn and thus the reverse
turn expands at the expense of the longer helix. Finally the lengths
of the two helices making up the hairpin are approximately equal in
length. During the process of creeping, which lasts for $1.6\cdot
10^5$ MC scans, the energy drops by 20~kcal/mol to about
$-1190$~kcal/mol. This drop is mainly due to the formation of X-X
pairs, rising in number from one to about five. As a further
result of the process of creeping, $R_{ends}$ falls from 15~{\AA},
shortly after coalescence, to 4--5~{\AA} (Figs.~\ref{fig:combi1}(c) and
\ref{fig:combi2}(c)), i. e. the hairpin is complete.

After the creeping process has stopped and a compact helical hairpin
conformation is reached, the structural fluctuations of the hairpin
are reduced considerably. In the picture of the folding funnel
\cite{Wol1995} the trajectories have reached a thermodynamic
bottleneck, where the multiple folding pathways approach the native
state from different sides and are slowed down by entropic
barriers. Nevertheless further rearrangements can be observed. For
example in trajectory (1) after $1.2\cdot 10^6$ MC scans there is a
cooperative reorganization of the two termini which goes along with an
increase of helix content by one residue, and an increase in the
number of X-X pairs to a value fluctuating between five and six
(Fig.~\ref{fig:sdg}(b)). Due to these changes the energy goes down
from $-1190$~kcal/mol to about $-1200$~kcal/mol.  In trajectory (2) a
remarkable development takes place at $10^6$ MC scans
(Fig.~\ref{fig:sdg}(c)) when the glycines in the turn region, which
previously had flickered between helical and other turn conformations,
are forming a short $\alpha$-helix which is stable for a few thousand
MC scans. This process is also visible in the energy trace as a
transient increase by about 10~kcal/mol. At the same time the number
of hydrophobic X-X pairs decreases momentarily by two. The glycine
helix collapses very suddenly, and all but one glycine are finally
stabilized in a non-helical turn conformation. After the collapse of
the glycine helix, energy and number of X-X pairs return to their
previous levels, but the conformation has changed to a near-native one
with a stable non-helical turn of glycines between two helices.

The folding pathways in trajectories (1) and (2) are consistent with
the diffusion-collision model (DCM) of Karplus and Weaver
\cite{Kar1976} which provides a theoretical framework for protein
folding and is supported by a large number of experimental findings
\cite{Kar1994}. As pointed out in the introduction, the essence of the
DCM is that in the folding process microdomains are formed,
e. g. $\alpha$-helices, which diffuse relatively to each
other.  
Eventually they collide and then coalesce with a certain
probability. This behaviour is observed in trajectories (1) and
(2). The helices are formed and diffuse relatively to each other as
quasi rigid bodies by random changes of the interhelical angle. In the
DCM, a collision of microdomains does not necessarily imply that they
remain together. This has been observed in trajectory (1) where after
a first collision the two helices separate again and continue their
relative diffusion. The DCM further allows that ``after partial
collapse and/or weak coalescence of microdomains to a more compact
structure with a non-native conformation, the attainment of the native
conformation might involve surface diffusion in one or two
dimensions'' \cite{Kar1994}. The creeping motion observed in both
trajectories is such a one-dimensional diffusion. It is also
reminiscent of the reptation movement introduced by de Gennes
\cite{deG1971} for polymers in polymer melts and hence could be called
self-reptation. Note that this self-reptation
movement is different from other dynamic processes where a net
increase of helix content is observed in parallel to a formation of a
helix dimer (e.g. as in \cite{Vie1994}). During self-reptation of the
present $\alpha$-helical hairpin the total helix content remains
approximately constant and helical turns are shifted from the
Glycine stretch to the neighbouring amino acids.

The DCM not only pictures the folding process qualitatively but also
allows the prediction of measurable quantities. For example it is
possible to estimate the time $\tau_f$ for the folding of the two
helices, i. e. the time between their generation and their ultimate
coalescence, using Eq.~\ref{eq:tauf} \cite{Bas1988a,Kar1994}:
\begin{equation}
\label{eq:tauf}
\tau_f=l^2/D + L \Delta V(1-\beta)/(D A \beta),
\end{equation}
where $l$ is a length related to the size of the diffusion space, $D$
is the relative diffusion coefficient of the microdomains, $L$ is a
length related to folding/unfolding rates and also to the size of the
diffusion space, $\Delta V$ is the diffusion volume, $A$ is the
collision surface area, and $\beta$ is the relative coalescence
probability or sticking probability for a collision event. Following
the procedures for the evaluation of these quantities in
Refs.~\cite{Bas1988a,Kar1994} and assuming that the microdomains are
two helices of twelve and eight residues, respectively, connected by a
loop of six residues, we obtain the following values:
$l^2=664$~{\AA}$^2$, $D=0.138$~{\AA}$^2/$ps, $L=16.3$~{\AA}, $\Delta
V=2.30\cdot 10^5$~{\AA}$^3$, and $A=2590$~{\AA}$^2$. In
Refs.~\cite{Bas1988a,Kar1994} $\beta$ is treated as a free
parameter. A large number of simulations of the type presented in this
section could be used to determine the probability $\beta$ more
accurately. For a very crude first estimate of
$\beta$ we restrict ourselves to the current data. Since there a
near-coalescence occurred only once, the value of $\beta$ should be of
the order of one for the model polypeptide. A value of 0.5 seems a
reasonable guess for an order of magnitude estimation. Assuming these
parameter values for the quantities in Eq.~\ref{eq:tauf} we find
$\tau_f=2.29\cdot 10^{-8}$~s. This time can be used to estimate the
time corresponding to one MC scan. An inspection of the trajectories
shows that the helices coalesce after a diffusion period of about
$2\cdot 10^5$ MC scans. Thus one MC scan is approximately equal to
0.1~ps, and hence each of the MC trajectories runs over about
160~ns. Independently and based on a comparison of dynamic MC with MD
simulations over long times we have recently estimated the MC scan to
be of the same order of magnitude \cite{Hoffmann1996b}, i.~e. about
two orders of magnitude larger than the time step of conventional
MD. Earlier estimates based on comparisons with the Rouse polymer
model \cite{Rou1953} yielded a value for the time corresponding to one
MC scan which was one order of magnitude larger but did not consider
contributions of non-bonded interactions \cite{Kna1992a}.

Previously it was thought that because of the long times involved in
the diffusion process, simulations at the atomic level could not be
carried out long enough to show aspects of the DCM during protein
folding. Therefore simulations have been restricted to the microdomain
level of resolution using preformed and explicitely stabilized
microdomains \cite{Lee1987,Bas1988a,Kar1994}. These simulations were
valuable to explore general questions concerning the DCM in larger
proteins, but of course under these simulation conditions it had to be
expected that the trajectories would obey the DCM. Other workers have
used simplified lattice models \cite{Sik1990}; \cite{Sko1991b} and
found no DCM behaviour. Instead in these models ``rather, the helices
that form native hairpins are constructed on-site, with folding
initiating at or near the turn'' \cite{Sik1990}. It had been suspected
that the disagreement with DCM may be due to the local moves that had
been employed in these lattice simulations which would not allow the
diffusion of intact microdomains \cite{Kar1994}. Our results show that
this argument is not valid in this general form, because the
trajectories described above clearly show the diffusion of
microdomains despite the use of local moves. It seems more likely that
the disagreement with DCM is related to the combined use of lattice
models and local moves. Interestingly, Rey and
Skolnick \cite{Sko1991} compared lattice simulations and off-lattice
Brownian Dynamics simulations. Whereas in the lattice simulations the
folding of an $\alpha$-helical hairpin was not consistent with a DCM
like process, some of the Brownian Dynamics trajectories clearly
followed the DCM scheme.

In the case of our model polypeptide the existence
of relatively stable helices certainly promotes the folding according
to the DCM. But relatively stable and fast folding polyalanine-based
helices are not unusual \cite{Mar1989,Williams1996}, hence for the
given sequence of amino acids, which is dominated by alanines, one
should expect DCM like folding. Experimentally, for arbitrary
sequences mixtures of various folding mechanisms are observed,
including global diffusion of larger parts of the respective proteins
(see e.g. Ref.~\cite{Ballew1996}).

	\section*{Conclusions}
Trajectories of a small model protein were produced using a dynamic MC
method with window moves. These off-lattice MC moves generate
efficiently and realistically conformational changes of a
polypeptide. A molecular model in atomic detail with only dihedral
angle degrees of freedom, and an energy function derived from a
conventional MD model were employed. Due to the ability of MC to
generate larger conformational changes per move than MD can generate
per step of time propagation, the MC method reaches longer time
regimes than MD with the same amount of CPU-time.

Starting from an extended $\beta$-strand conformation
or from loop conformations
of the protein glutamine synthetase, four out of six
trajectories equilibrated into a native-like $\alpha$-helical hairpin
conformation within about $10^6$ MC scans. Three phases of time
evolution can be clearly distinguished in these trajectories:
1. fast formation of two $\alpha$-helices separated by a turn,
2. diffusion of the interhelical angle, 3. coalescence of the two
helices followed by a self-reptation into a native-like
helix-turn-helix conformation. All three phases are consistent with
experiments and in accord with predictions made by the DCM
\cite{Kar1994}.

The comparison of the folding time predicted for the $\alpha$-helical
hairpin by the DCM with the folding times observed in the MC
trajectories yielded a value for the time corresponding to one MC scan
of the order of 0.1~ps. This value is in agreement with an
independent estimate using a comparison of MD and dynamic MC
simulations \cite{Hoffmann1996b}.

Further work is necessary to allow a quantitative comparison of such
simulations with experimental results. In particular the energy
function has to be refined to consider more quantitatively
contributions of the solvent and of a greater variety of sidechains.

\subsection*{Acknowledgment}
The authors gratefully acknowledge the help of Fredo Sartori with the
force field. The CHARMM code has been provided by Prof. Martin Karplus
and Molecular Simulations Inc. This work is supported by the European
Union contract ERBCHRXCT930112, the Deutsche Forschungsgemeinschaft,
project Kn329/1, and by the Fonds der Deutschen Chemischen Industrie.

	\subsubsection*{Figure captions}

\begin{figure}[h]
        \caption[]{\baselineskip4ex
Conformations from the two trajectories starting from two loop
conformations of polypeptide chain F of the protein glutamine
synthetase (2glsF). White and black spheres are C$_{\alpha}$ atoms of
glycines and hydrophobic X residues, respectively. The figures were
generated using Molscript \cite{Kra1991}. N-terminus of polypeptide
 is always the upper end. Upper left: start conformation
corresponding to loop 13 to 39 of 2glsF. Lower left: conformation of
lowest energy (-1219~kcal/mol) in trajectory starting from
conformation in upper left (after $2.8\cdot 10^6$ MC scans). All-atom root mean
square deviation (RMSD) to reference structure is 3.2~{\AA}. Upper
right: start conformation corresponding to loop 163-189 of 2glsF. Lower
right: conformation of lowest energy (-1212~kcal/mol) in trajectory
starting from conformation in upper right (after $1.6\cdot 10^6$ MC
scans). RMSD to reference structure is 2.7~{\AA}.}
\label{fig:loopStarts}
\end{figure}
\vspace{1ex}

\begin{figure}[h]
        \caption[]{\baselineskip4ex
Traces of energy, number of pairs of hydrophobic X residues, end-end
distance $R_{ends}$, and interhelical angle in MC trajectory (1).  The
abscissa of part (d) gives the time in units of $10^5$ MC scans. It is
used for all four parts of this figur. The value of the respective
quantities after every 5000th MC scan is depicted. (a) Conformational
energy. (b) Number of pairs of X residues. Two X residues are
considered a pair if they are not neighbours in sequence and their
C$_\beta$-atoms are closer than 6~{\AA}. (c) Distance $R_{ends}$
between first and last $C_\alpha$-atom of the polypeptide chain. (d)
Interhelical angle defined as angle between sum of $C=O$ bondvectors
in the first helix (residues 2 to 6), and sum of $O=C$ bondvectors in
the second helix (residues 15 to 22).}
\label{fig:combi1}
\end{figure}
\vspace{1ex}

\begin{figure}[h]
        \caption[]{\baselineskip4ex
Trajectory (2). Axes are the same as in Fig.~\ref{fig:combi1}. Note
however that in part (d) the interhelical angle refers to the
angle between the sum of $C=O$ bondvectors of residues 4 to 11, and
sum of $O=C$ bondvectors of residues 19 to 24.}
\label{fig:combi2}
\end{figure}

\vspace{1ex}
\begin{figure}[h]
        \caption[]{\baselineskip4ex
Time evolution of ($\phi,\psi$) values of each of the 26 residues
(structure-dynamograms). (a) Graylevel code for structure-dynamograms
(parts (b) and (c)). Each graylevel codes for a rectangular region in the
Ramachandran plot around a secondary structure, e. g. white codes for
$\alpha_R$-helical residues. Residues having no canonical structure
are coded in black (here not shown for technical reasons). Parts (b)
and (c) are structure-dynamograms of MC trajectories (1) and (2),
respectively. Conformations are depicted after every $10^4$th MC scan.
}
\label{fig:sdg}
\end{figure}

\vspace{1ex}
\begin{figure}[h]
        \caption[]{\baselineskip4ex
Coalescence and self reptation in MC trajectory
(1). The time is given in numbers of MC scans. C$_\alpha$-atoms of
glycines and hydrophobic X residues are shown as white and black
spheres, respectively. Wide ribbons are $\alpha$-helix turns. N and C
indicate N- and C-terminus, respectively. The pictures where generated
using Molscript \cite{Kra1991}. 
}
\label{fig:creeping1}
\end{figure}

\vspace{1ex}
\begin{figure}[h]
        \caption[]{\baselineskip4ex
Coalescence and self reptation in trajectory (2). See also caption of
Fig.~\ref{fig:creeping1}.
}
\label{fig:creeping2}
\end{figure}
\pagebreak
\clearpage

	\clearpage
\thispagestyle{empty}
Fig.~\ref{fig:loopStarts} of Hoffmann and Knapp
\vspace{4cm}

\begin{figure}[h]
        \leavevmode
        \centering
\psfig{file=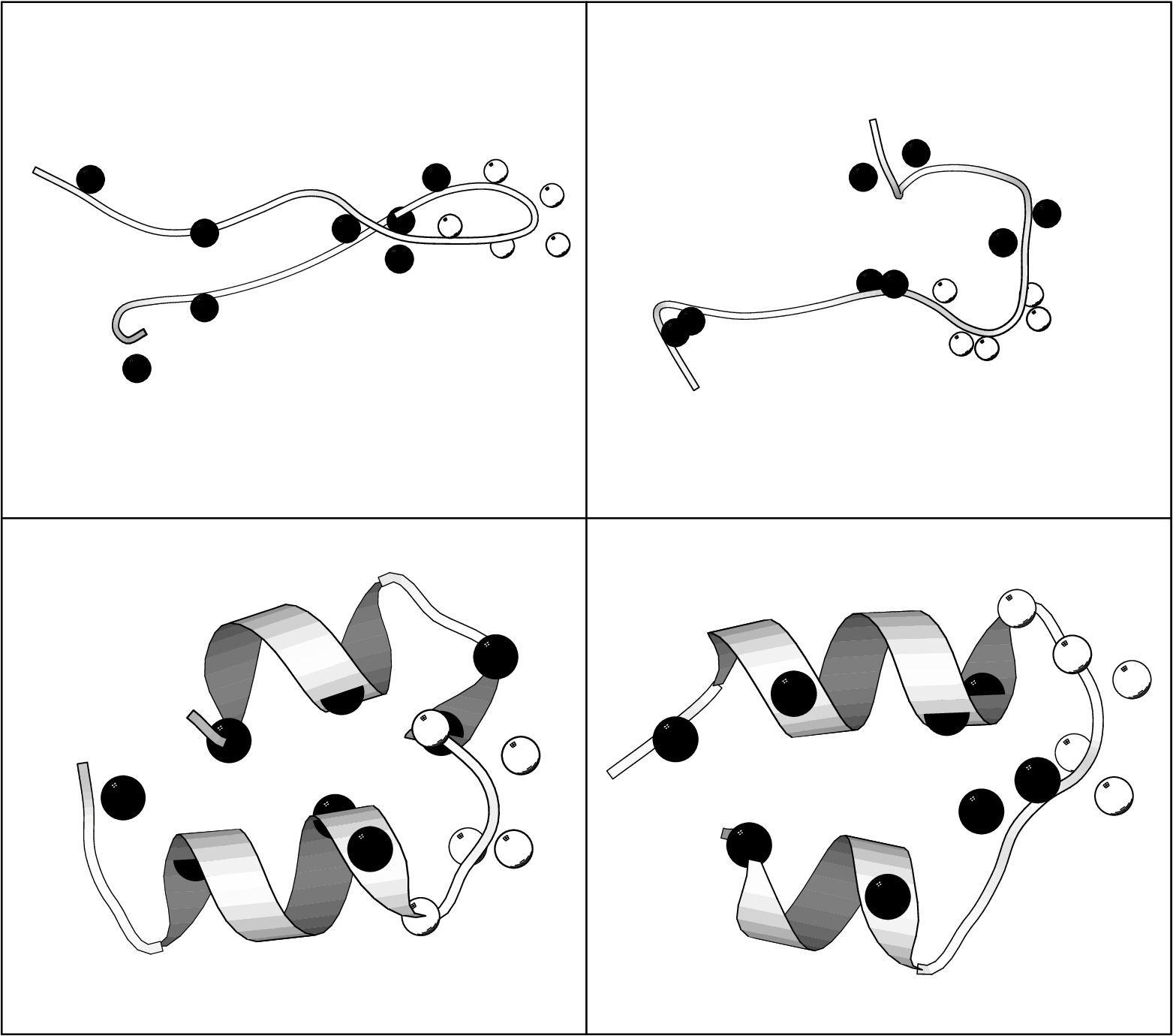,bbllx=39pt,bblly=49pt,bburx=721pt,bbury=651pt,width=80mm}
\end{figure}

\clearpage
Fig.~\ref{fig:combi1} of Hoffmann and Knapp
\vspace{4cm}

\begin{figure}[h]
        \leavevmode
        \centering
\psfig{file=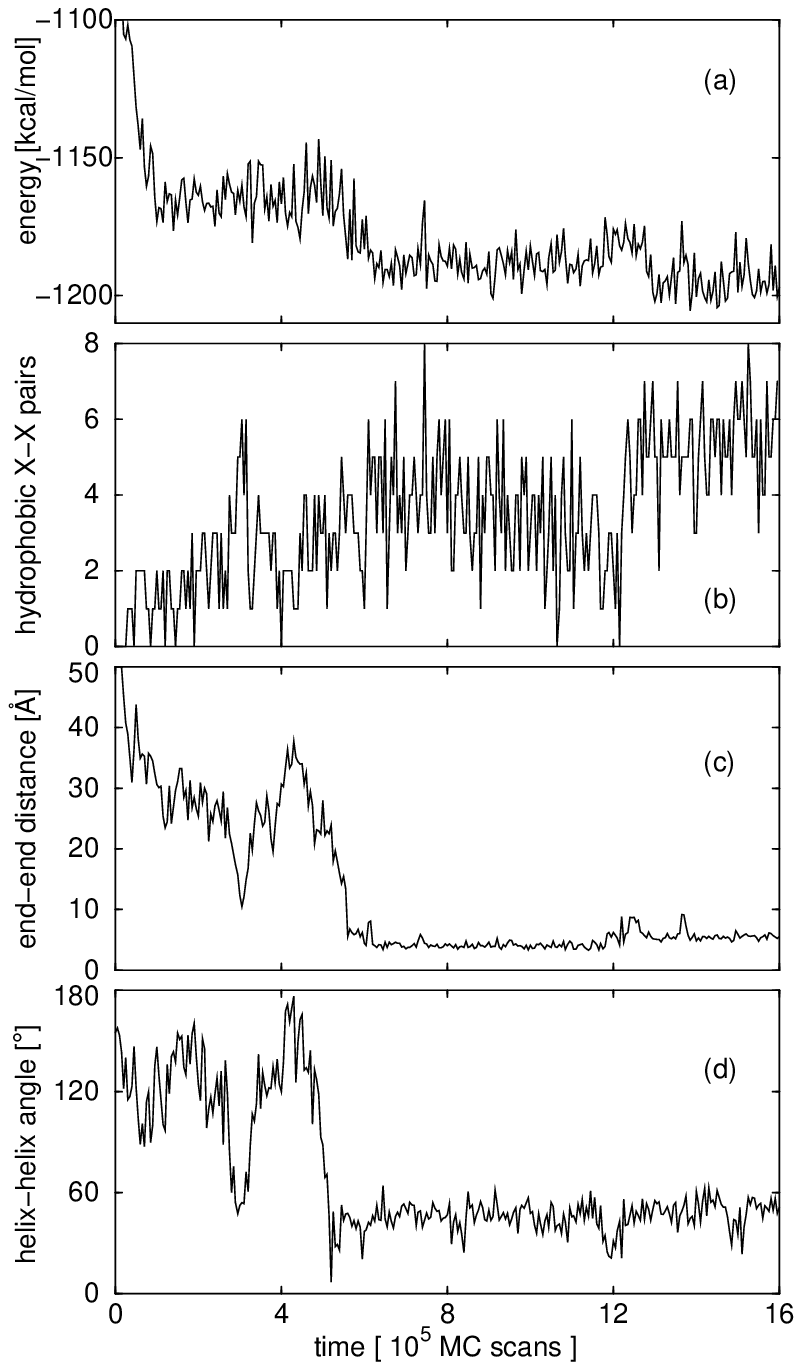,bbllx=178pt,bblly=341pt,bburx=406pt,bbury=741pt,width=80mm}
\end{figure}

\clearpage
\thispagestyle{empty}
Fig.~\ref{fig:combi2} of Hoffmann and Knapp
\vspace{4cm}

\begin{figure}[h]
        \leavevmode
        \centering
\psfig{file=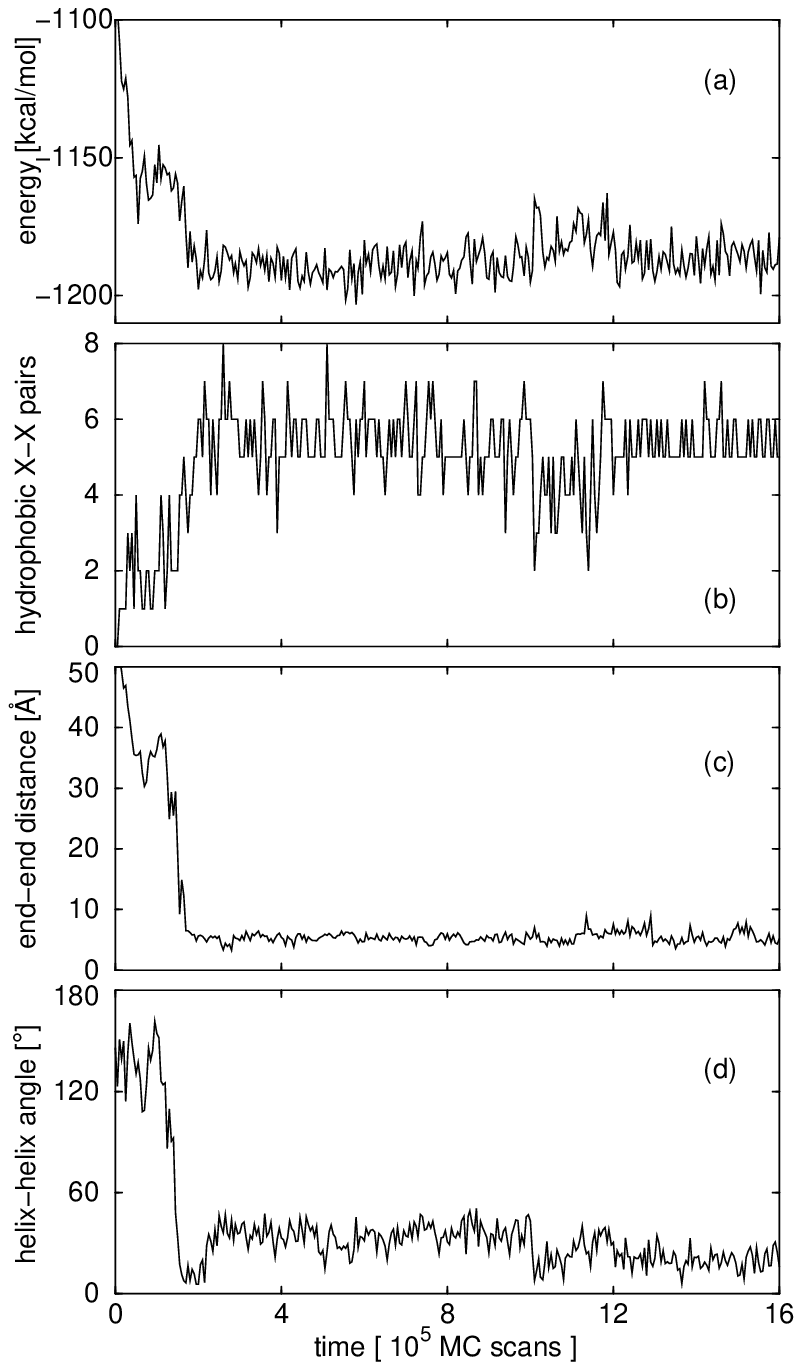,bbllx=178pt,bblly=341pt,bburx=406pt,bbury=741pt,width=80mm}
\end{figure}

\clearpage
\thispagestyle{empty}
Fig.~\ref{fig:sdg} of Hoffmann and Knapp
\vspace{4cm}

\begin{figure}[h]
        \leavevmode
        \centering
\psfig{file=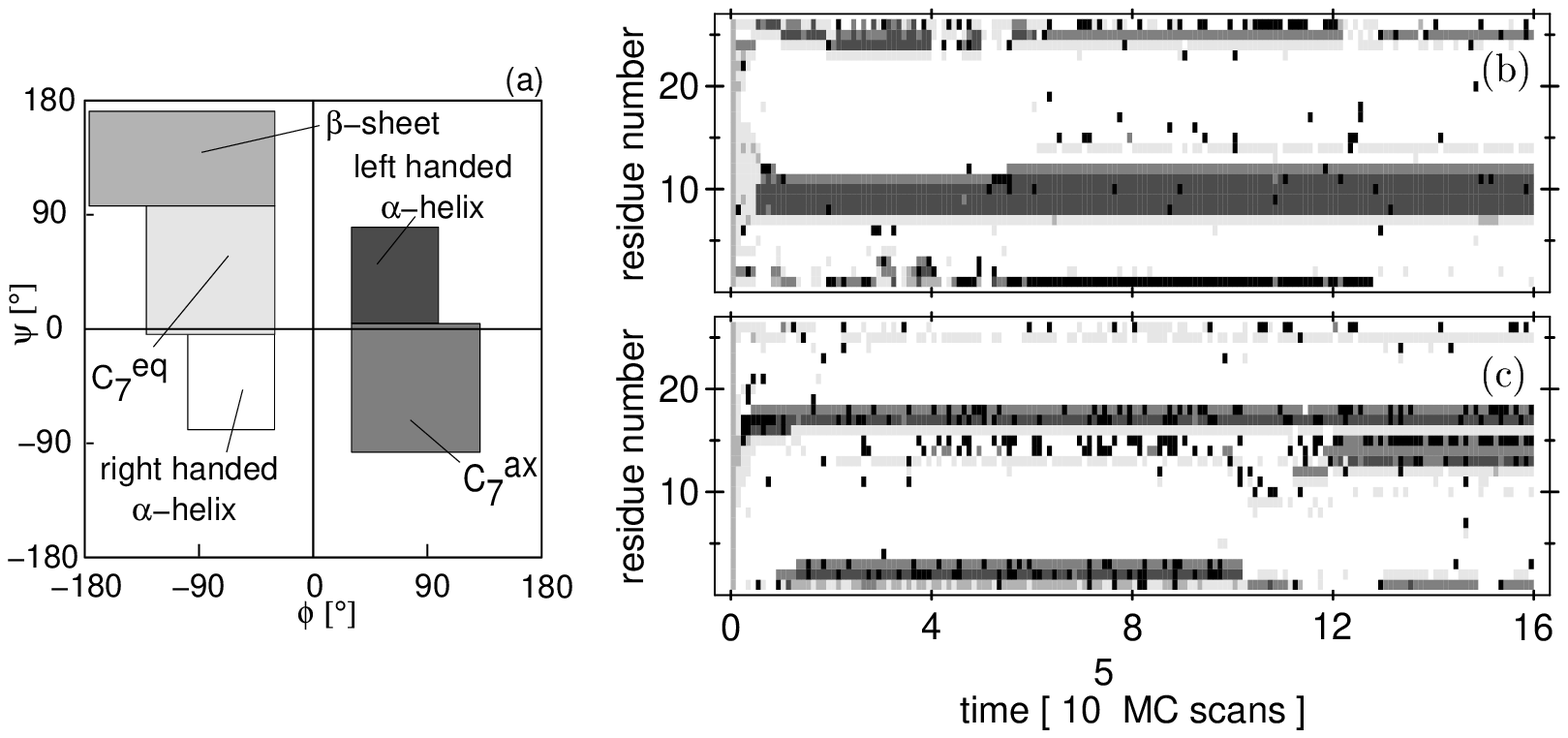,bbllx=73pt,bblly=371pt,bburx=544pt,bbury=587pt,width=165mm}
\end{figure}

\clearpage
\thispagestyle{empty}
Fig.~\ref{fig:creeping1} of Hoffmann and Knapp
\vspace{4cm}

\begin{figure}[h]
        \leavevmode
        \centering
\psfig{file=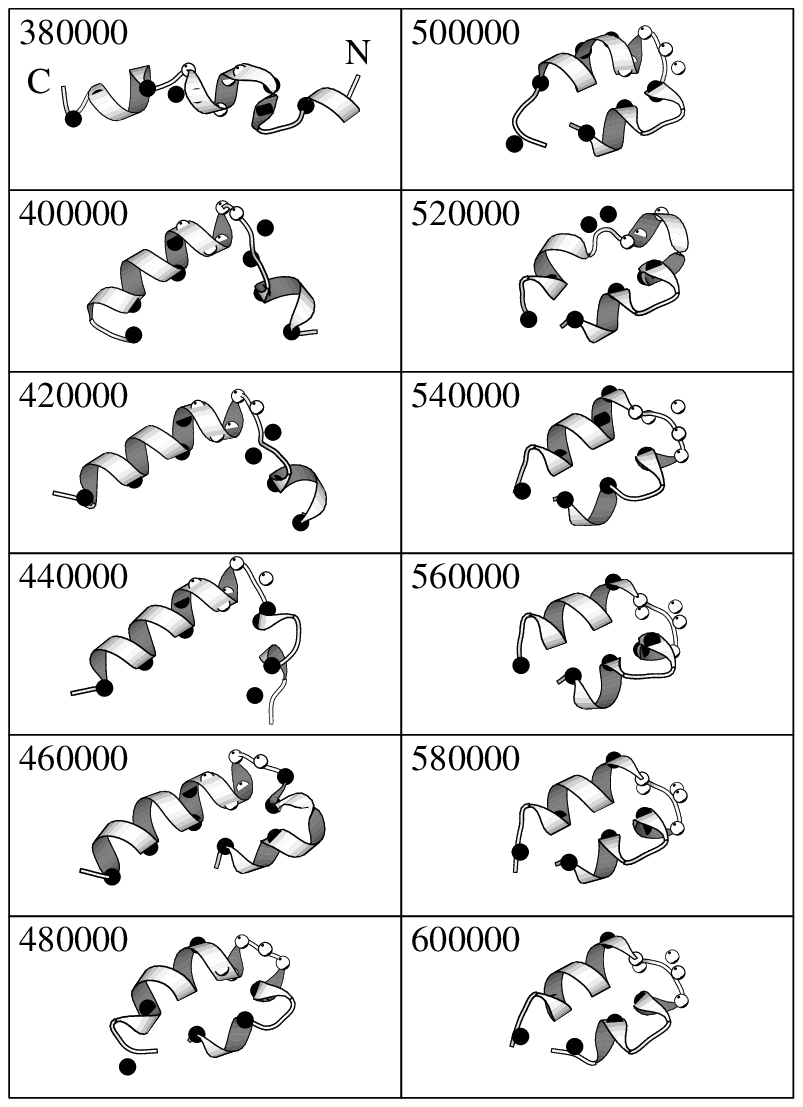,bbllx=177pt,bblly=418pt,bburx=406pt,bbury=738pt,width=80mm}
\end{figure}

\clearpage
\thispagestyle{empty}
Fig.~\ref{fig:creeping2} of Hoffmann and Knapp
\vspace{4cm}

\begin{figure}[h]
        \leavevmode
        \centering
\psfig{file=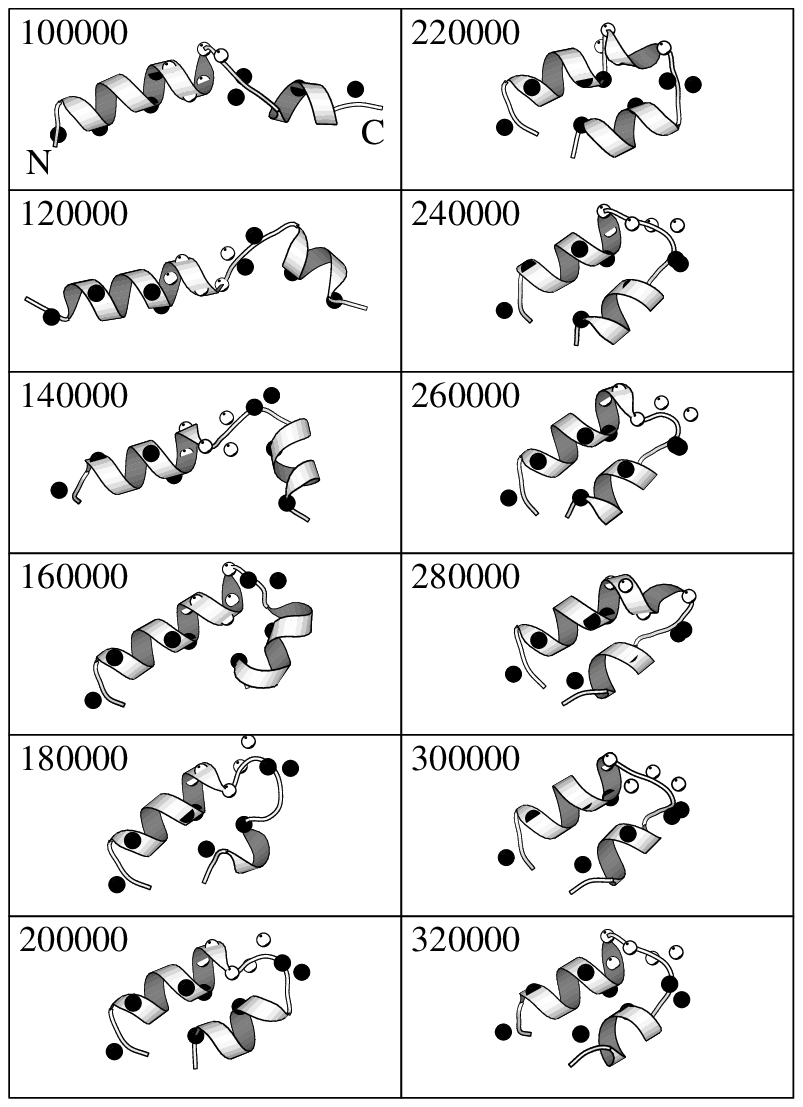,bbllx=177pt,bblly=418pt,bburx=406pt,bbury=738pt,width=80mm}
\end{figure}

\end{document}